\documentclass[prl,twocolumn,superscriptaddress,nofootinbib]{revtex4-2}

\usepackage{amsmath,amssymb,amsfonts,mathtools,bm}
\usepackage{physics}
\usepackage{siunitx}
\usepackage{graphicx}
\usepackage[colorlinks=true,linkcolor=blue,citecolor=blue,urlcolor=blue]{hyperref}

\newcommand{\ii}{\mathrm{i}}
\renewcommand{\dd}{\mathrm{d}}
\renewcommand{\tr}{\operatorname{tr}}
\renewcommand{\bf}[1]{\mathbf{#1}}

\newcommand{\Gr}{\mathrm{Gr}}
\newcommand{\BZ}{\mathrm{BZ}}

\begin{document}

\title{Relaxation toward an Ideal Chern Band through Coupling to a Markovian Bath}

\author{Bruno Mera}
\affiliation{Instituto de Telecomunica\c{c}\~oes and Departmento de Matem\'{a}tica, Instituto Superior T\'ecnico, Universidade de Lisboa, Avenida Rovisco Pais 1, 1049-001 Lisboa, Portugal}
\affiliation{Advanced Institute for Materials Research (WPI-AIMR), Tohoku University, Sendai 980-8577, Japan}

\author{Tomoki Ozawa}
\affiliation{Advanced Institute for Materials Research (WPI-AIMR), Tohoku University, Sendai 980-8577, Japan}

\begin{abstract}
We propose a microscopic, weak-coupling mechanism by which generic Chern bands asymptotically relax toward ideal bands. We consider coupling interacting electrons to a Caldeira-Leggett-like Ohmic bosonic bath. Using the Born-Markov approximation, we analytically show that, upon taking the leading order contribution in momentum of the form factor, Slater determinant states of a Chern band under Hartree-Fock approximation evolve toward Slater determinant states corresponding to an ideal Chern band. We also numerically validate our proposal by performing simulation of a massive Dirac model with the extended Hubbard interaction, showing that the Berry curvature and quantum metric co-evolve toward saturation of the trace condition.
Our proposal provides a concrete dissipative mechanism for driving Chern bands toward ideal quantum geometry, a fundamental building block for the stabilization of fractional Chern insulators.
\end{abstract}

\maketitle
\paragraph{Introduction---} The notion of \emph{ideal Chern bands}~\cite{rahul:2014,claassen:2015,ozawa:mera:2021,mera:ozawa:2021:relations,mera:ozawa:2021:engineering,wang:cano:millis:liu:yang:2021,ledwith:vishwanath:parker:2023,estienne:regnault:valentin:2023,mera:ozawa:2023}---flat bands whose Berry curvature and quantum metric saturate a quantum-geometric inequality---has emerged as a central paradigm in the theory of fractional Chern insulators (FCIs)~\cite{bernevig:2011,titus:2011,sheng:2011,sun:2011,wen:2011} and topological flat-band materials~\cite{wang:cano:millis:liu:yang:2021,ledwith:tarnopolsky:khalaf:vishwanath:2020, wang:liu:2022,ledwith:vishwanath:parker:2023,ledwith:vishwanath:khalaf:2022} due to their remarkable similarity to the lowest Landau level. More recently, a generalization of this notion---\emph{generalized Landau levels} or \emph{harmonic bands}~\cite{liu:mera:fujimoto:ozawa:wang:2025,fujimoto:parker:dong:khalaf:vishwanath:ledwith:2025,paiva:wang:ozawa:mera:2025,onishi:avdoshkin:fu:2025}---was formulated, and their suitability for stabilizing non-Abelian FCI phases was numerically confirmed.

Although the importance of the ideal Chern bands and generalized Landau levels have been increasingly appreciated, how one can obtain such bands, either theoretically or experimentally, remains to be nontrivial. Theoretically, one needs special models which fulfill the ideal conditions. Experimentally, there is no known simple method to obtain ideal Chern bands or generalized Landau levels.
In this paper, we propose a generic mechanism under which ideal Chern bands or generalized Landau levels can be obtained. We show that, by coupling weakly interacting electrons to Caldeira-Leggett type Markovian bath, mean-field (Slater determinant) states evolve into states which minimize the Dirichlet energy functional determined from the quantum geometry. When the Dirichlet energy is dominant over the kinetic energy, the resulting state one obtains in a long-time limit is nothing but the ideal Chern band. Numerical simulations for the massive Dirac model confirm this evolution, with the Dirichlet energy approaching the topological lower bound $\pi|\mathcal{C}|$, where $\mathcal{C}$ is the Chern number of the occupied states, as the Berry curvature and quantum metric coevolve toward the ideal-band condition. In this picture, ideal Chern bands correspond to the global minima of the Dirichlet energy, while generalized Landau levels are stationary points.
We note that recently, in Refs.~\cite{yu:lian:ryu:2025,fukui:2025}, gradient flow with respect to the Dirichlet energy of the Bloch projector was proposed to achieve the ideal-band conditions. Compared to these recent works, our work provides a concrete dissipative mechanism through which ideal bands can be achieved. 
In this work, the bosonic environment is modeled by a Caldeira–Leggett bath, viewed as a generic effective description of dissipation rather than as a microscopic model for a specific material platform. Such an effective bath can arise, for instance, from couplings to phonons or to electromagnetic modes, such as those associated with external radiation fields or cavity fields, after expanding the relevant fields in modes and making approximations such as the rotating-wave approximation.
\paragraph{Model and projection to Slater dynamics---}  We first consider, for demonstration, a single-site model and later extend it to a lattice model. The model consists of a system of interacting electrons coupled to a Caldeira-Legget-like bosonic bath,
$\widehat H=\widehat H_S+\widehat H_B+\widehat H_I$,
where
\begin{align}
\widehat H_S&=\sum_{i,j}t_{ij}\,c_i^\dagger c_j
+\tfrac12\sum_{i,j,k,l}v_{ijkl}\,c_i^\dagger c_j^\dagger c_l c_k, \nonumber\\
\widehat H_B&=\sum_\alpha \Omega_\alpha\, b_\alpha^\dagger b_\alpha,\quad
\widehat H_I=\lambda\sum_{i,j,\alpha}
\left(b_\alpha^\dagger \;c^\dagger_{i}\left(V_\alpha\right)_{ij}c_j +\text{H.c.}\right).
\label{eq: microscopic model}
\end{align}
Above, $c_i^\dagger$ and $c_i$ create and annihilate an electron in orbital \(i\),
while \(b_\alpha^\dagger\) and \(b_\alpha\) create and annihilate a bosonic bath mode \(\alpha\) of frequency \(\Omega_\alpha\).
The $t_{ij}$ encode single-particle hoppings, and
$v_{ijkl}$ denote the two-body interaction matrix elements.
The system–bath coupling is parameterized by a real constant $\lambda$ and by the matrices $V_\alpha$,
whose elements $(V_\alpha)_{ij}$ specify the interaction vertex between orbital $i$ and $j$
mediated by bath mode $\alpha$. We note that the system-bath Hamiltonian preserves the system's charge.

Assuming (i) an initial Slater-determinant state and Gaussian fermionic density matrices at all times---so that Wick’s theorem holds---(ii) a weak-coupling Born factorization
$\widehat\rho(t)\simeq \widehat\rho_S(t)\otimes\widehat\rho_B(t)$,
and (iii) a Markov approximation, we can show that the system remains to be in a Slater determinant state during the dynamics. We provide the full proof of this nontrivial fact in Supplemental Material~\cite{sm}.

Since under our assumptions the system remains in the Slater determinant state, we from now on consider only the Slater determinant states. The Slater determinant states can be mathematically represented by a point on a Grassmannian manifold, which we denote as $\Gr$, with the rank equal to the particle number described by the single-particle density matrix $P=[P_{ij}]=[\langle c_j^\dagger c_i\rangle]$.

In the absence of system-bath coupling ($\widehat H_I=0$),
the dynamics reduces to the time-dependent Hartree-Fock equation
\begin{equation}
\ii\frac{\dd P}{\dd t }=[T+V(P),P],\qquad
V_{ik}(P)=\sum_{jl}\!\big(v_{ijkl}-v_{ijlk}\big)P_{lj},
\label{eq:tdHF}
\end{equation}
where $T=[t_{ij}]$ is the hopping matrix and $V(P)=[V_{ij}(P)]$,
with $V_{ij}(P)=\sum_{kl}\!\big(v_{ikjl}-v_{iklj}\big)P_{lk}$ the mean-field interaction potential.
This evolution can also be written in the following form
\begin{equation}
\frac{\dd P}{\dd t}=\ii\,\Bigl[P,\frac{\partial H_S}{\partial P}\Bigr], \label{eq:hamflow}
\end{equation}
where the Hamiltonian function $H_S=H_S(P)=\tr\left[\left(T+\frac{1}{2}V(P)\right)P\right]$ is the expectation value of the system Hamiltonian $\widehat H_S$ evaluated on a Slater determinant, and $\frac{\partial H_S}{\partial P}=[\frac{\partial H_S}{\partial P_{ij}}]$ denotes the derivative with respect to $P$,
which coincides with the single-particle Hartree-Fock Hamiltonian $\frac{\partial H_S}{\partial P}=T+V(P)$.
(We note that Eq.~(\ref{eq:hamflow}) has a geometrical meaning as a Hamiltonian flow on the Grassmannian; see SM~\cite{sm} for further discussion.)

\paragraph{Ohmic bath leads to metriplectic flow---}
Solving the bath equations formally using retarded Green’s functions and inserting the result into the system’s equation of motion leads to a memory kernel governed by the bath spectral density
\begin{align}
J(\omega)=\sum_\alpha \delta(\omega-\Omega_\alpha)\ket{V_\alpha}\bra{V_\alpha},
\end{align}
which is a superoperator, and in this context it means a linear map on the vector space of matrices in which $P(t)$ takes values. We assume that $J(\omega)$ takes the Ohmic form
\begin{align}
J(\omega)=\eta\,\omega e^{-\omega/\omega_c}\, \mathbb{I},
\end{align}
where $\mathbb{I}$ is the identity superoperator and $\eta$ is the overall dissipation rate. In the Markovian approximation, which includes $\omega_c\to\infty$, memory effects are negligible and the dynamics becomes local in time, yielding the equation (see SM~\cite{sm} for a detailed derivation)
\begin{equation}
\frac{\dd P}{\dd t}
= \ii\,\Bigl[P,\frac{\partial H_S}{\partial P}\Bigr]
- \gamma\,\Bigl[P,\Bigl[P,\frac{\partial H_S}{\partial P}\Bigr]\Bigr],
\qquad
\gamma = \lambda^2 \eta > 0.
\label{eq: metriplectic}
\end{equation}
This equation has a geometrical meaning of \emph{metriplectic} dynamics on the Grassmannian, which is a combination of conservative Hamiltonian dynamics and an irreversible disspative dynamics; see SM~\cite{sm} for further discussion.
\paragraph{Monotonically decreasing energy along the flow---}
Consider the system's Hamiltonian function $H_S(P)$.
Using the identity $\Pi_P(X)=[P,[P,X]]=(1-P)XP+PX(1-P)$,
which gives the orthogonal projection of an Hermitian operator $X$ onto the tangent space
$T_P\mathrm{Gr}$ with respect to the Hilbert-Schmidt inner product,
one finds
\begin{equation}
\frac{\dd H_S}{\dd t}
=\mathrm{tr}\!\left[\frac{\partial H_S}{\partial P}
\frac{\dd P}{\dd t} \right]
=-\gamma\,\mathrm{tr}\!\left[\Pi_P\Bigl(\frac{\partial H_S}{\partial P}\Bigr)^2\right]\le 0,
\label{eq: Lyapunov}
\end{equation}
and, hence, $H_S(P)$ is nonincreasing along trajectories. In the longtime limit, $H_S(P(t))$ should then approach the lowest value, and hence $\frac{\dd H_S}{\dd t} = 0$ as $t \to \infty$. This implies $\Pi_P\Bigl(\frac{\partial H_S}{\partial P}\Bigr) = 0$ in the longtime limit, which one can show is equivalent to $[P,\frac{\partial H_S}{\partial P}]=0$. We call the set determined by $[P,\frac{\partial H_S}{\partial P}]=0$ to be the \emph{Hartree-Fock critical set}, which is the set of states one obtains in the longtime limit. (For a more rigorous derivation of these steps; see Supplemental Material~\cite{sm}.)

\paragraph{Generalization to lattice models and emergence of the Dirichlet energy---}
We can generalize the microscopic model of Eq.~\eqref{eq: microscopic model}
to a translation-invariant two-dimensional lattice with $N$ internal degrees of freedom per site
(e.g., orbitals, spin, or sublattice).
Now the annihilation and creation operators in Eq.~\eqref{eq: microscopic model} should carry positional indices $\mathbf{r}$ as well as indices for internal degrees of freedom. After performing Fourier transformation to move to momentum space, we denote the Bloch Hamiltonian describing the hopping to be
$T(\mathbf{k})=[t_{ij}(\mathbf{k})]$
and short-range density-density interactions to be
controlled by a potential $v(\mathbf{q})$. The system Hamiltonian in momentum space then reads
\begin{equation}
\widehat{H}_S=\lambda_T\sum_{\mathbf{k}}\sum_{i,j=1}^{N}
c_{\mathbf{k},i}^\dagger t_{ij}(\mathbf{k})c_{\mathbf{k},j}
+\frac{1}{V}\sum_{\mathbf{q}} v(\mathbf{q})\,{:n_{-\mathbf{q}}n_{\mathbf{q}}:}.
\end{equation}
Here $n_{\mathbf{q}}$ denotes the Fourier transform of the density operator,
and $\lambda_T$ sets the overall hopping strength. We restrict attention here to interactions coupling through the total density operator $n_{\mathbf q}$. In more general multiorbital settings, the interaction kernel may carry orbital indices, in which case the  geometric term in $H_S$ can depend on the orbital embedding and need not be controlled solely by the usual band quantum metric, as obtained below in this simplified setting. The quantum metric itself, as noted independently by Simon and Rudner~\cite{simon:rudner:2020} and Haldane~\cite{haldane:2005}, also depends on the choice of orbital embedding. This choice must be made consistently with the physical interaction, since changing the relative orbital positions modifies not only the quantum metric but also the momentum-space interaction kernel whenever the density-density potential depends on orbital separations. Assuming the system occupies a Slater determinant obtained by filling a Chern band,
described by an $N\times N$ orthogonal rank-$r$ projector
$P(\mathbf{k})=[P_{ij}(\mathbf{k})]$, and following the same Born-Markov approximation used for the model in Eq.~\eqref{eq: microscopic model} together with an Ohmic-bath assumption homogeneous in momentum (see SM~\cite{sm}), the evolution of the band is governed by a generalization of Eq.~\eqref{eq: metriplectic},
\begin{equation}
\frac{\partial P(\mathbf{k})}{\partial t}
=\ii\!\left[P(\mathbf{k}),\frac{\delta H_S}{\delta P(\mathbf{k})}\right]
-\gamma\!\left[P(\mathbf{k}),
\left[P(\mathbf{k}),\frac{\delta H_S}{\delta P(\mathbf{k})}\right]\right],
\label{eq: metriplectic-flow-Chern}
\end{equation}
where $\frac{\delta H_S}{\delta P(\mathbf{k})}
=[\frac{\delta H_S}{\delta P_{ij}(\mathbf{k})}]$ denotes the functional derivative of the Hartree-Fock energy density functional
\begin{align}
H_S
&=\lambda_T\!\int_{\BZ^2}\!\!\frac{\dd^2\mathbf{k}}{(2\pi)^2}
\,\mathrm{tr}\!\big[T(\mathbf{k})P(\mathbf{k})\big]
\nonumber\\[-2pt]
&\quad
-\!\!\int_{\BZ^2}\!\!\int_{\BZ^2}\!\!
\frac{\dd^2\mathbf{k}\,\dd^2\mathbf{q}}{(2\pi)^4}\,
v(\mathbf{q})\,\mathrm{tr}\!\big[P(\mathbf{k})P(\mathbf{k}-\mathbf{q})\big]. \label{eq:HS}
\end{align}
The appearance of functional derivatives is because we have a continuum of components $P_{ij}(\mathbf{k})$ parametrized by $\mathbf{k}$.
For a finite lattice with (twisted or not) periodic boundary conditions, this functional derivative naturally reduces to an ordinary derivative, since only a finite set of momenta appears in the Fourier decomposition of $P$. We now expand the form factor, $\mathrm{tr}\!\big[P(\mathbf{k})P(\mathbf{k}-\mathbf{q})\big]$ in Eq.~(\ref{eq:HS}) in small $\mathbf{q}$ and take the lowest nontrivial part.
(We will later consider the cases without the small $\mathbf{q}$ approximation.)
The lowest order $\mathbf{q}^0$ term just gives an overall constant so we ignore this term in the discussion below, and the first order term vanishes. Expanding up to a quadratic order in $\mathbf{q}$, we write
\begin{align}
H_S
&\simeq\lambda_T\!\int_{\BZ^2}\!\!\frac{\dd^2\mathbf{k}}{(2\pi)^2}
\,\mathrm{tr}\!\big[T(\mathbf{k})P(\mathbf{k})\big] + \frac{2\lambda_{\mathrm D}}{(2\pi)^2}\,E_{\mathrm D}[P], \label{eq:hsapprox}
\end{align}
where $\lambda_{\mathrm D} \equiv (1/4\pi)\int_{0}^{Q}\! \dd q\, v(q)\,q^{3}$, with $Q$ being the momentum cutoff, and
\begin{align}
E_{\mathrm D}(P)
\equiv \frac{1}{4}\!\int_{\BZ^2}\!\!\dd^2\mathbf{k}\,\sqrt{\det h}\,h^{ij}\,
\mathrm{tr}\!\!\left(
\frac{\partial P}{\partial k^i}\frac{\partial P}{\partial k^j}
\right),
\end{align}
with $h$ determined through the anisotropy of the interaction $v(\mathbf{q}) = v(\sqrt{h_{ij}q^i qj})$.
We assume that the anisotropy of $v(\mathbf{q})$ comes from the anisotropy of the underlying lattice, in which case one can show that $h_{ij}$ is equal to the Brillouin-zone–integrated quantum metric,
\begin{align}
h_{ij}\propto \int_{\BZ^2}\!\frac{\dd^2\mathbf{k}}{(2\pi)^2}\, g_{ij}(\mathbf{k}), \label{eq:hij}
\end{align}
and $h^{ij}$ is its inverse.
See Supplemental Material~\cite{sm} for the detailed derivation of Eqs.~(\ref{eq:hsapprox}-\ref{eq:hij}). The functional derivative of $H_S$ is then
$(2\pi)^2\frac{\delta H_S}{\delta P(\mathbf{k})}=\lambda_T T(\mathbf{k})-\lambda_{\mathrm D}\Delta_h P(\mathbf{k}),$
where $\Delta_h P=h^{ij}\frac{\partial^2 P}{\partial k^i\partial k^j}$
is the anisotropic Laplacian associated with $h$.

We call the first term in Eq.~(\ref{eq:hsapprox}) as \textit{kinetic term}, and the second term the \textit{Dirichlet energy term}. The Dirichlet energy is the integral of the trace of the quantum metric~\cite{paiva:wang:ozawa:mera:2025}—$E_{\mathrm{D}}(P)=(1/4)\int_{\BZ^2}d^2\bf{k}\;\tr_{h}(g)$—and, due to the trace inequality $\tr_h(g)\geq |F_{12}|$, where $F_{12}$ is the Berry curvature, it thus satisfies the topological bound $E_{\mathrm{D}}\geq\pi|\mathcal{C}|$. We are interested in the regime where the Dirichlet energy term dominates over the kinetic term, i.e. $\lambda_{\mathrm D} \gg \lambda_T$ so that the dynamics is dominated by the Dirichlet energy. 

\paragraph{Relaxation toward ideal bands---} The Dirichlet energy functional has extrema given by \emph{generalized Landau levels} (nonzero Chern number harmonic maps)~\cite{liu:mera:fujimoto:ozawa:wang:2025,paiva:wang:ozawa:mera:2025},
which satisfy $Q(\Delta_h P)P=0$, equivalently $[P,\Delta_h P]=0$.
\emph{Ideal bands} are precisely the minima of $E_{\mathrm D}$;
they saturate the topological bound $E_{\mathrm D}\ge \pi|\mathcal{C}|$,
the global version of the quantum-geometric trace inequality
relating the Berry curvature and quantum metric.
In the limit $\lambda_{\mathrm D}\gg\lambda_T$,
the Hartree-Fock energy $H_S$ is dominated by the Dirichlet term.
In this regime, $E_{\mathrm D}$ decreases monotonically
under the flow of Eq.~\eqref{eq: metriplectic-flow-Chern},
while the Chern number $\mathcal{C}$ is conserved.
The last statement is true because the Chern class of the band depends only on the homotopy class of $P_t:\BZ^2\to\Gr$, and this homotopy class is preserved by the flow, due to continuity. In contrast, as we will discuss below, when the small-$\mathbf q$ approximation breaks down, the map $P_t$ can develop a discontinuity, giving rise to a topological phase transition in which the Chern number drops; see Fig.~2 and SM~\cite{sm} for a comparison between the two situations.

\paragraph{Numerical demonstration: Massive Dirac model---}
We integrate Eq.~\eqref{eq: metriplectic-flow-Chern} for a two-band Dirac model (see, for instance, Ref.~\cite{mera:goldman:zhang:2022} for more details on the quantum geometry of this model),
$T(\mathbf{k})=\vec{d}(\bf{k})\cdot\vec{\sigma}$ with 
$\vec{d}(\bf{k})=(\sin k_x,\sin k_y,M-\cos k_x-\cos k_y)$, $M=-0.5$ so the lower band has $\mathcal{C}=1$. Writing $P(t,\bf{k})=(I+\vec{n}(t,\bf{k})\cdot\vec{\sigma})/2$, the flow becomes a Landau-Lifshitz-Gilbert equation
\begin{equation}
\frac{\partial\vec{n}}{\partial t}
=-\,\vec{n}\times\!\vec{h}
+\gamma\,\vec{n}\times\!\left(\vec{n}\times\!\vec{h}\right), \quad (2\pi)^2\vec{h}=2\lambda_T\vec{d}-\lambda_{\mathrm{D}}\nabla^2\vec{n},
\label{eq: LLG equation}
\end{equation} 
with initial condition $\vec{n}(t=0,\bf{k})=-\vec{d}(\bf{k})/|\vec{d}(\bf{k})|$.

\begin{figure}[h!]
 \includegraphics[width=0.88\columnwidth]{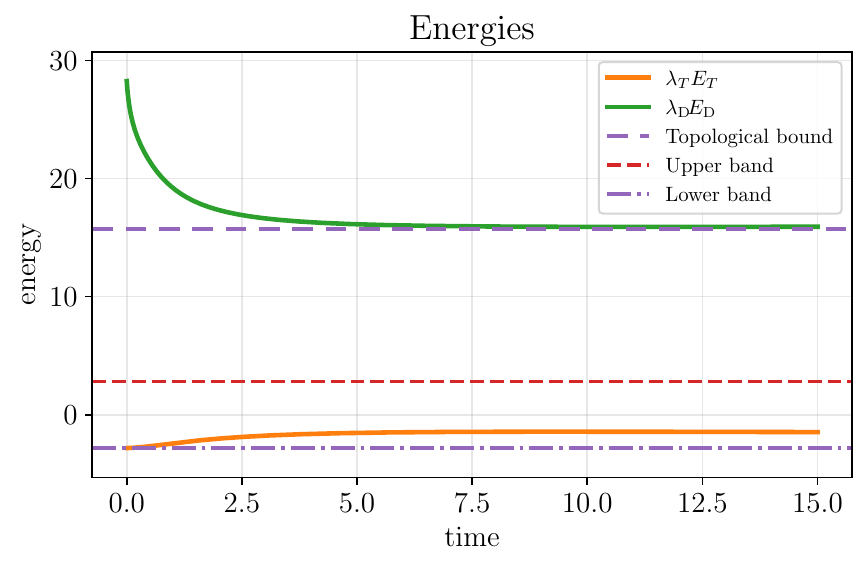}
  \caption{
    Time evolution of the Dirichlet and hopping energy components during the time evolution according to Eq.~\eqref{eq: LLG equation}.
    The Dirichlet energy monotonically decreases and approaches the topological lower bound
    $E_{\mathrm D}\ge\pi|\mathcal{C}|$, while the hopping energy $\lambda_T E_T$, with $E_T=\int_{\BZ^2}\frac{\dd ^2\bf{k}}{(2\pi)^2}\vec{d}(\bf{k})\cdot \vec{n}(t,\bf{k})$, remains nearly constant equal to that of the lowest band. The lower and upper band energies are also plotted for reference.
  }
  \label{fig:energy}
\end{figure}

Using $(\gamma,\lambda_{\mathrm{D}},\lambda_T)=(1.5,1.25,0.025)$, we show in Fig.~\ref{fig:energy} the energy as a function of time and confirm that the Dirichlet energy $E_{\mathrm{D}}(t)$ decreases monotonically to $\pi \mathcal{C}=\pi$, which is the value one should achieve for the ideal Chern bands.
In Fig.~\ref{fig:energy}, we compare the Dirichlet energy $E_{\mathrm{D}}(t)$ and the hopping energy $E_{\mathrm{T}}(t)$; we are in a regime where the Dirichlet energy is dominant over the hopping energy, which is the condition we required in order to have the flow evolve toward the ideal Chern bands. See SM~\cite{sm} for further numerical results on the saturation of the trace inequality.

\begin{figure}[t]
\centering
\includegraphics[width=0.88\columnwidth]{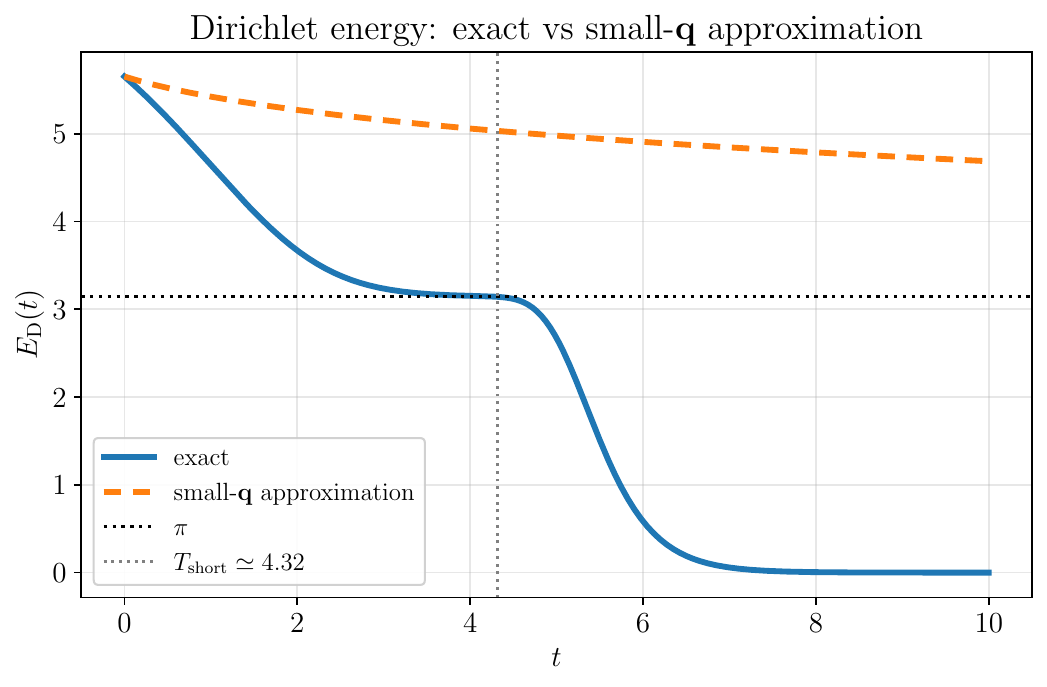}
\caption{Dirichlet energy for the exact evolution and the small-$\mathbf q$ approximation for the extended Hubbard interaction. The exact
evolution comes closest to the ideal-band value $E_{\mathrm D}=\pi|\mathcal C|$ at $T_{\mathrm{short}}\simeq 4.32$, while the
band remains in the topological sector $\mathcal C=1$; see SM~\cite{sm} for the corresponding plot of $\mathcal C$. We take the parameters $(U,V,\lambda_T,Q)=(8.0,0.75,0.025,0.5\pi)$, the corresponding Dirichlet coupling is found to be $\lambda_{\mathrm D}\simeq 1.183$ (see SM~\cite{sm}).
}
\label{fig:extended_hubbard_energy}
\end{figure}

\begin{figure*}[t]
\centering
\includegraphics[width=1.76\columnwidth]{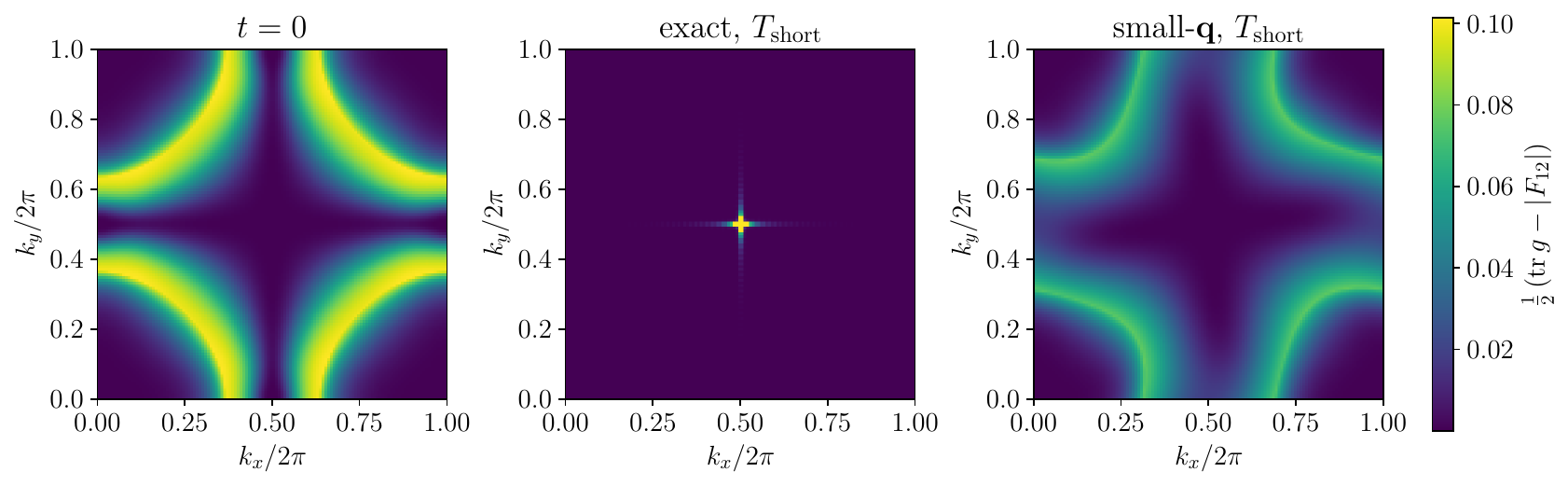}
\caption{Deviation from saturation of the trace inequality for the extended Hubbard interaction at $t=0$ (left panel) and $t=T_{\mathrm{short}}\simeq 4.32$ for the exact evolution (center panel) and the evolution under the small-$\bf{q}$-approximation (right panel).
}
\label{fig:extended_hubbard_saturation}
\end{figure*}

To see the effect of the small $\mathbf{q}$ approximation of the form factor $\mathrm{tr}\!\big[P(\mathbf{k})P(\mathbf{k}-\mathbf{q})\big]$, we have also numerically studied the evolution
Eq.~\eqref{eq: metriplectic-flow-Chern} for the exact Hartree-Fock energy functional in Eq.~\eqref{eq:HS}, using an extended Hubbard interaction
$v(\mathbf q)=U+2V(\cos q_x+\cos q_y)$, where $U$ describes an onsite interspecies Hubbard interaction and $V$ a nearest-neighbor interaction; see the Supp. Mat. for details. We compare this exact evolution with the evolution with small-$\mathbf q$ approximation. The results are presented in Figs.~\ref{fig:extended_hubbard_energy} and~\ref{fig:extended_hubbard_saturation}.

As shown in Fig.~\ref{fig:extended_hubbard_energy}, we find that even the exact evolution, without the small $\mathbf{q}$ approximation, drives the initially generic Chern band close to an ideal band. The Dirichlet energy reaches its closest to the topological bound $\pi |\mathcal C|=\pi$ at $t=T_{\mathrm{short}}$, before the subsequent change of topological class. This behavior is further confirmed by Fig.~\ref{fig:extended_hubbard_saturation},
which shows that, at $T_{\mathrm{short}}$, the exact evolution nearly saturates the trace inequality throughout the Brillouin zone, except in a small
neighborhood of $\bf{k}=(\pi,\pi)$. Figs.~\ref{fig:extended_hubbard_energy} and~\ref{fig:extended_hubbard_saturation} show that the exact evolution reaches this near-ideal regime
more rapidly; we have confirmed that this tendency holds for a wide range of parameters as long as the onsite interaction $U$ dominates over the nearest-neighbor interaction $V$. (See SM~\cite{sm} for more details.)
We find that, for the exact evolution, after reaching the near-ideal Chern band, the system goes through a topological phase transition and become a trivial insulator with $\mathcal{C} = 0$, eventually reaching a trivial insulator with no quantum geometric structure.
In SM~\cite{sm}, we discuss this transition in terms of a bubbling mechanism and further investigate the robustness of the relaxation toward ideal-band geometry as a function of the interaction parameters $(U,V)$. These results provide a direct numerical check that the relaxation mechanism toward ideal quantum geometry is not an artifact of the small-$\bf{q}$ approximation, but holds more generically. 

\paragraph{Discussion and outlook. ---}
Although we could give analytical discussions only for the cases with small $\mathbf{q}$ approximation, our numerical results indicate that the relaxation toward ideal Chern band is a more generic feature without the need for the small $\mathbf{q}$ approximation. It would be of interest to further systematically study the cases beyond small $\mathbf{q}$ approximation to understand the general mechanism of the the relaxation.

How our methods apply to models of the strongly correlated regime relevant to fractional Chern insulators will be further discussed in future works. Generalized Landau levels, on the other hand, correspond to extrema of the energy functional. We expect that, once the flow reaches an extremum, the system remains to be in the generalized Landau level for a certain period of time, before further relaxing toward the ideal Chern band. Further exploration of precise engineering of generalized Landau levels using our method will be discussed in future works. Once the ideal Chern bands and generalized Landau levels are obtained, with further addition of appropriate short-range quantum inter-particle interactions, the ground state of the system should exhibit fractional quantum Hall effects. Our proposal thus opens an avenue toward obtaining fractional Chern insulator phases starting from generic Chern bands.

\emph{Acknowledgments---}B.~M. acknowledges support from the Security and Quantum Information Group (SQIG) in Instituto de Telecomunica\c{c}\~{o}es, Lisbon. This work is funded by FCT/MECI through national funds and when applicable co-funded EU funds under UID/50008: Instituto de Telecomunicações (IT). B.~M. further acknowledges the Scientific Employment Stimulus --- Individual Call (CEEC Individual) --- 2022.05522.CEECIND/CP1716/CT0001, with DOI: \href{https://doi.org/10.54499/2022.05522.CEECIND/CP1716/CT0001}{10.54499/2022.05522.CEECIND/CP1716/CT0001}. T.~O. acknowledges support from JSPS KAKENHI Grant Number JP24K00548, JST PRESTO Grant No. JPMJPR2353.

\bibliography{bib.bib}
\end{document}